# ONTOLOGY BASED APPROACH FOR VIDEO TRANSMISSION OVER THE NETWORK

Rachit Mohan Garg[1], Yamini Sood[2], Neha Tyagi[3]

[1]Deptt. of Computer Science & Engineering, Jaypee University of Information Technology, Waknaghat, Solan, H.P, India
`rachit.mohan.garg@gmail.com`
[2]Deptt. of Computer Science & Engineering, Jaypee University of Information Technology, Waknaghat, Solan, H.P, India
`eryaminisood@gmail.com`
[3]Deptt. of Computer Science & Engineering, Jaypee University of Information Technology, Waknaghat, Solan, H.P, India
`tyagi.neha14@gmail.com`

## ABSTRACT

*With the increase in the bandwidth & the transmission speed over the internet, transmission of multimedia objects like video, audio, images has become an easier work. In this paper we provide an approach that can be useful for transmission of video objects over the internet without much fuzz. The approach provides a ontology based framework that is used to establish an automatic deployment of video transmission system. Further the video is compressed using the structural flow mechanism that uses the wavelet principle for compression of video frames. Finally the video transmission algorithm known as RRDBFSF algorithm is provided that makes use of the concept of restrictive flooding to avoid redundancy thereby increasing the efficiency.*

## 1. INTRODUCTION

With the increase in the bandwidth & the transmission speed over the internet, transmission of multimedia objects like video, audio, images has become an easier work. The accelerated development of internet and the high diversity of networked devices available in the market have led to the fast development of networked multimedia applications (video transmission). These applications are usually statically deployed which encounters a lot of problem. To minimize the problems encountered during transmission of the multimedia objects such as video an approach is proposed in this paper.

The approach can be useful for transmission of video objects over the internet without much fuzz. The approach provides an ontology based framework that is used to establish an automatic deployment of video transmission system. In order to characterize such type of application, the MODA ontology's have been introduce to describe the communication tasks emerged during video transmission. Further the video is compressed using the structural flow mechanism that uses the wavelet principle for compression of video frames. Finally the video transmission algorithm known as RRDBFSF algorithm is provided that makes use of the concept of restrictive flooding to avoid redundancy thereby increasing the efficiency. By knowing the information of neighbor node in finite scope, this algorithm does breadth first search and selects the least number of forwarding neighbor nodes to reduce redundant information in broadcasting routing information.

In this paper section 2 presents MODA framework for the video transmission over the network. Section 3 discusses a compression technique for the video based on wavelet coding. Section 4 provides a transmission protocol for transmitting the video over to the intended receiver. The paper is concluded in section 4 followed by the references.

## 2. RELATED WORK [12][13][19]

### 2.1. H.261/H.263

The H.261 and H.263 are not International Standards but only Recommendations of the ITU. They are both based on the same technique as the MPEG standards and can be seen as simplified versions of MPEG video compression. They were originally designed for video-conferencing over telephone lines, i.e. low bandwidth. However, it is a bit contradictory that they lack some of the more advanced MPEG techniques to really provide efficient bandwidth use.

The conclusion is therefore that H.261 and H.263 are not suitable for usage in general digital video coding.

### 2.2. MPEG-1

In Motion JPEG/Motion JPEG 2000 each picture in the sequence is coded as a separate unique picture resulting in the same sequence as the original one. In MPEG video only the new parts of the video sequence is included together with information of the moving parts.

MPEG-1 is focused on bit-streams of about 1.5 Mbps and originally for storage of digital video on CDs. The focus is on compression ratio rather than picture quality. It can be considered as traditional VCR quality but digital instead. Only the decoder is actually standardized. An MPEG encoder can be implemented in different way and a vendor may choose to implement only a subset of the syntax, providing it provides a bit stream that is compliant with the standard. This allows for optimization of the technology and for reducing complexity in implementations. However, it also means that there are no guarantees for quality.

### 2.3. MPEG-3

The next version of the MPEG standard, MPEG-3 was designed to handle HDTV, however, it was discovered that the MPEG-2 standard could be slightly modified and then achieve the same results as the planned MPEG-3 standard. Consequently, the work on MPEG-3 was discontinued.

### 2.4. MPEG-4

The next generation of MPEG, MPEG-4, is based upon the same technique as MPEG-1 and MPEG-2. Once again, the new standard focused on new applications. The most important new features of MPEG-4, ISO/IEC 14496, concerning video compression are the support of even lower bandwidth consuming applications, e.g. mobile devices like cell phones, and on the other hand applications with extremely high quality and almost unlimited bandwidth. In general the MPEG-4 standard is a lot wider than the previous standards. It also allows for any frame rate, while MPEG-2 was locked to 25 frames per second in PAL and 30 frames per second in NTSC. When "MPEG-4," is mentioned in surveillance applications today it is usually MPEG-4 part 2 that is referred to. This is the "classic" MPEG-4 video streaming standard, a.k.a. MPEG-4 Visual. Some network video streaming systems specify support for "MPEG-4 short header," which is an H.263 video stream encapsulated with MPEG-4 video stream headers. MPEG-4 short header does not take advantage of any of the additional tools specified in the MPEG-4 standard, which gives a lower quality level than both MPEG-2 and MPEG-4 at a given bit-rate.

## 2.3. MPEG-3

The next version of the MPEG standard, MPEG-3 was designed to handle HDTV, however, it was discovered that the MPEG-2 standard could be slightly modified and then achieve the same

## 3. PROPOSED APPROACH

### 3.1. MODA Framework for Video Transmission

The MODA process (CIM to PIM) allows generating from the sending communication task a SCA domain including both server and client composites. For each composite the SessionController and MediaController components are inferred. Following the general MODA process the adequate session controller as well as the media controller implementations has been selected. XSL templates used to implement the mapping rules the MODA engines. This is shown in figure 1.

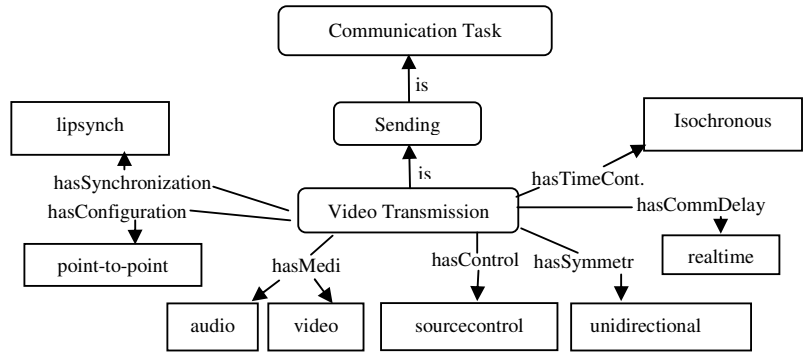

Figure 1. Video Transmission Process with MODA Framework

### 3.2. Compressing Video for Optimal Transmission [9]

In wavelet video coding, a *group of frames* (*gof*) is decomposed along the three major axes: temporal, horizontal and vertical. However, this decomposition does not take the regularity of the *gof* into account. In the presence of global motion, uniform 3D paths of regularity are defined in a *gof*, which extend along the direction of motion. The situation gets more complicated when the motion is a mixture of the local and global components. In this case, *subgroups of frames* (*subgofs*) with different motion types result in multiple directions of regularity. One way of modelling this regularity is modelling the motion. The pixel correspondence information over multiple frames gives the directions of regularity of the *gof*. The motion-compensated (MC) wavelet coding algorithms use this approach.

The wavelet decomposition is applied to the *gof* by partitioning it into *subgof* so that directions of regularity of each *subgof* are as closely estimated as possible. This can be done by minimizing the compression cost of each subgroup of frames $F_i$ so that the total cost becomes

$$D + \lambda R = \sum_i D_i + \lambda R_i \qquad (1)$$

where $D_i$ is the sum of squared reconstruction error of $F_i$, $R_i$ is the bit cost of the wavelet and flow coefficients, and $\lambda$ is a Lagrange multiplier. This is depicted in equation 1. This segmentation is achieved by partitioning the *gof* into rectangular prisms known as cuboids using an *oct tree*. The width of each dimension of a cuboid is $2^{k_j}$, where $j \in \{1; 2; 3\}$ denotes the particular dimension. The wavelet coefficients are quantized using the quantization parameter, $\Delta$, and then are encoded. Since the bit cost of these coefficients is almost proportional to the number of non-zero coefficients, as shown in [8], the bit cost of the wavelet coefficients can be approximated as $R_{b,i} = \beta_0 M_i$, where $M_i$ is the number of non-zero coefficients and $\beta_0 = 7$.

The choice of $\lambda$ as a function of the quantization parameter $\Delta$ can be computed by minimizing the total cost equation with respect to $\Delta$. This minimization results in the definition of $\lambda$ as $\lambda = 3\Delta^2/4\beta_0$. The minimization of the total cost starts with computing the cost of all cuboids in the *oct tree*. The cost, $(D_i + \lambda R_i)$, can be minimum for only one of the four flow classes, including the no-flow case. In the end, the flow class that has the minimum cost determines the flow type of $F_i$.

The optimal segmentation of $F$ is found by a split/merge algorithm starting from the leaf nodes of the *oct tree*. At each level, eight child nodes are merged into a single node if their cumulative cost is greater than the parent's cost, otherwise they stay split. The split-merge algorithm is applied until the top of the tree is reached, which concludes the segmentation of the *gof* in terms of the bit rate and the reconstruction error. The basis for the whole *gof* is called the *block orthonormal wavelet basis*, and it consists of the union of the bases of the *subgof*s in the final segmentation, on their own supports.

### 3.3. Radius Restrained Distributed Breadth First Search Flooding Algorithm (RRDBFSF) [10][11] for Video Transmission

This algorithm is flooding in a small scope, namely, radius restrained flooding algorithm, and can reduce redundancies within a certain scope. Based on the rule that lessen the cost and time of forwarding message between nodes to the best, we choose the scope within a radius of three to flood message. So, every node need know its neighbor nodes which are connected directly with it, and need realize some information about their neighbor nodes. We call these information are nodes information within a radius of three. Thus, we suppose that x is random node in the networks.

### 3.3.1. Description of RRDBFSF Algorithm

#### 3.3.1.1. Rules Defining

The general rule is that the radius is three in the algorithm. Therefore, there are some defines of rules:

a) The least cost of forwarding message time;
b) Node is only concerned about the nodes flooding within a radius of three;
c) Always choose the shortest path.

#### 3.3.1.2. Given Conditions of RRDBFSF Algorithm

a) The network is connected entirely.
b) Any connection between nodes is bidirectional.
c) Before running the flooding algorithm, the node has received its neighbor nodes information and built the neighbor nodes table NT (x).

#### 3.3.1.3. Information Collection Procedure

Some common terms are defined in table 1.

Table 1. Node Set which defines

| Notation | Description |
|---|---|
| N(x) | The neighbor nodes set of node x. |
| RN(x) | The relative neighbor nodes set of node x, viz. RN (x). |
| N (x) | The set is calculated by the RRDBFSF algorithm. |
| TLen(x) | The neighbor nodes set of node x within a radius of three. |
| RTLen(x) | The relative neighbor nodes set of node x within a radius of three, viz. RTLen (x). |
| T(x) | The sum of neighbor nodes and the nodes within a radius of three |
| R(x) | The set of node x next forwarding nodes |

Table 2. Neighbor Nodes Table – NT(x)

| Neighbor node ID | Its neighbor nodes within a radius of three |
|---|---|

NT (x) is founded on the messages from neighbor nodes and its content is flashed real time and RT (x) is forwarding nodes table. It includes the node x next forwarding neighbor nodes and their respective next forwarding neighbor nodes.

Table 3. Next Forwarding Nodes Table – RT(x)

| Next forwarding node ID | Next forwarding nodes |
|---|---|

The RRDBFSF algorithm calculates, gets RT(x), and transmits it to next forwarding neighbor nodes. So, if a forwarding node ID 'i' is in RT(x), its next forwarding nodes set concerned with node x is RT (x, i). We take node 'e' as example. The terms define above are illustrated on the basis of figure 2 below.

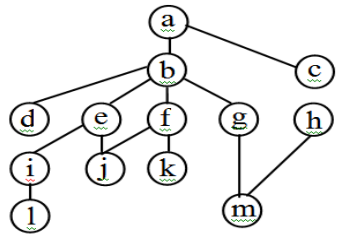

Figure 2. A Sample Communication Network

Sample of term defined above w.r.t node 'e':

N (e) = {b, i, j}

TLen (e) = {c, k, l, m, j, b}

T (e) = {b, i, j, c, k, l, m, b}

Table 4. Sample of Neighbor Nodes Table

| Neighbor Node ID | Neighbor Node in radius of 3 |
|---|---|
| b | f, g, h, l |
| i | f, a, g, d |

| j | a, g, d, f, e |
|---|---|

## 4. CONCLUSION AND FUTURE WORK

In this paper we present a framework for the video transmission & depict how by using Radius Restrained Breadth First Search we can decrease the redundant information in the network & stream our video data effectively to the destination node. In first part of the paper we present the MODA framework which is intended for automatic deployment and configuration guided by the user requirements and preferences. Then a compression technique to compress the video so as to transmit it effectively over the network is discussed. In the third part RRDBFSF algorithm is discussed, in which by realizing the information of neighbor nodes in finite scope i.e. 3, a BFS search is performed and the least redundant information i.e. route is selected.

The future work includes designing a more efficient algorithm which will route the video data efficiently & securely over the network. The other area of research is designing a framework for packet repairing which incurs less overhead on the basis of time & cost. Next steps in MODA framework aim at generating dynamic user interfaces for final users (to drive the deployment process) and enhancing deployment specifications produced by MODA in order to be used for self-configuring the communication system.

The International journal of Multimedia & Its Applications (IJMA) Vol.3, No.1, February 2011

**Authors**

**Rachit Mohan Garg**

The author is pursuing his Post Graduation from Jaypee University of Information Technology. He has completed his Engineering from Vishveshwarya Institute of Engineering & Technology; Ghaziabad affiliated to Gautam Buddh Technical University in 2008.

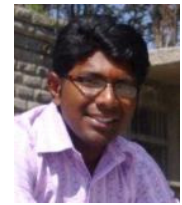

**Yamini Sood**

The author is pursuing her Post Graduation from Jaypee University of Information Technology with Graph Mining as her research field. She has completed her Engineering from Shri Sai College of Engineering & Technology; Badhani, Pathankot affiliated to Punjab Technical University in 2009.

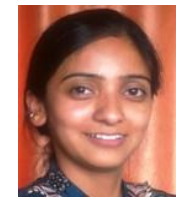

**Neha Tyagi**

The author is pursuing her Post Graduation from Jaypee University of Information Technology She has completed his Engineering from Dr. M.C.Saxena College of Engg. & Technology; Lucknow affiliated to Gautam Buddh Technical University in 2009.

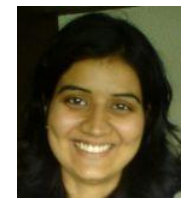